\documentclass[aps,pra,10pt,twocolumn,notitlepage,floatfix,
longbibliography]{revtex4-1}

\usepackage[dvips]{graphicx}
\usepackage{subfigure}
\usepackage{amsmath}
\usepackage{amssymb}
\usepackage{bm}
\usepackage{color}
\usepackage{hyperref}
\usepackage{multirow}
\usepackage[dvipsnames]{xcolor}% add text color capabilities
\usepackage[colorinlistoftodos]{todonotes}
\usepackage{changes}
\definechangesauthor[color=red]{PB}

\def\Fig{Fig.~}

\begin{document}
%%%%%%%%%%%%%%%%%%%%%%%%%%%%%%%%%%%%%%%%%%%%%%%%%%%%%%%%%%%%%%%%%%%%%%%%%%%%%%%%%%%%%%%%%%%%%%%%%

\title{All-Optical Bose-Einstein Condensates in Microgravity}

\author{G. Condon}
\author{M. Rabault}
\author{B. Barrett}
\author{L. Chichet}
\author{R. Arguel}
\author{H. Eneriz-Imaz}
\author{D. Naik}
\author{A. Bertoldi}
\author{B. Battelier}
\author{P. Bouyer}
\affiliation{LP2N, Laboratoire Photonique, Num\'{e}rique et Nanosciences, Universit\'{e} Bordeaux--IOGS--CNRS:UMR 5298, 1 rue Fran\c{c}ois Mitterrand, 33400 Talence, France}

\author{A. Landragin}
\affiliation{LNE-SYRTE, Observatoire de Paris, Universit\'{e} PSL, CNRS, Sorbonne Universit\'e,
61 avenue de l'Observatoire, 75014 Paris, France}

\date{\today}

%================================================================================================
% Outline:
% - Intro
% - présentation manip
% - production de condensats
% - simulateur
% - spectroscopie
% - conclusion
%================================================================================================

\begin{abstract}
We report on the all-optical production of Bose-Einstein condensates in microgravity using a combination of grey molasses cooling, light-shift engineering and optical trapping in a painted potential. Forced evaporative cooling in a 3-m high Einstein elevator results in $4 \times 10^4$ condensed atoms every 13.5 s, with a temperature as low as 35 nK. In this system, the atomic cloud can expand in weightlessness for up to 400 ms, paving the way for atom interferometry experiments with extended interrogation times and studies of ultra-cold matter physics at low energies on ground or in Space.
\end{abstract}

\maketitle

%================================================================================================
%\section{Introduction}
%\label{sec:Introduction}

The ability to produce ultra-cold matter and macroscopic quantum systems with laser cooling and evaporation techniques \cite{Dalibard1989, Ketterle1996} has brought about new physics predicted by quantum mechanics. It has also enabled new tests of fundamental physics by using matter waves for precision measurements. Today, Bose-Einstein condensates (BECs) \cite{Anderson1995} are employed as complex tools to test condensed matter phenomena, such as mimicking electronic solid state systems with atoms trapped in optical potentials \cite{Bloch2012}. Producing and controlling ultra-cold matter has also lead to ground-breaking methods for measuring inertial effects \cite{Barrett2014a}, time and frequency standards \cite{Ludlow2015}, and fundamental physical constants \cite{Clade2019} with high precision. This quantum toolbox opens exciting new prospects for inertial navigation \cite{Cheiney2018}, geodesy \cite{Trimeche2019}, tests of general relativity \cite{Overstreet2018}, and the detection of gravitational waves \cite{Canuel2018} and dark energy \cite{Elder2016}.

For all of these applications, gravity can be the ultimate limiting factor. The extremely low temperatures required to study some quantum phases of matter such as antiferromagnetism \cite{Mazurenko2017} are usually limited by gravitational sag. Temperatures of a few nK or below also correspond to very low energy and require long observation times hardly attainable under gravity. Finally, matter-wave interferometers require extended interrogation time \cite{LeCoq2006} to increase their sensitivity. Circumventing gravity can be achieved by using external forces to compensate the downward pull \cite{Leanhardt1513, Ricci2007} or to levitate the atoms \cite{Billy2008}. In each case, the additional magnetic or optical fields used to launch or hold the atoms against gravity results in residual perturbations. Moreover, when a magnetic field is used to support neutral atoms against the gravitational force, there is always a residual curvature which limits the adiabatic decompression of the trap \cite{Sackett2006}. Extended free-fall times can be achieved by launching atoms upward in an atomic fountain \cite{Kovachy2015}. Another method consists of letting the whole experiment fall freely under gravity, \emph{e.g.} in a drop tower \cite{VanZoest2010}, during parabolic flight in an aircraft \cite{Barrett2016} or a sounding rocket \cite{Becker2018}, or on the International Space Station \cite{Elliott2018}. Until now, BEC production on these platforms has been realized only with magnetic traps using atom chip technology.

\begin{figure}[!t]
  \centering
  \includegraphics[width=0.48\textwidth]{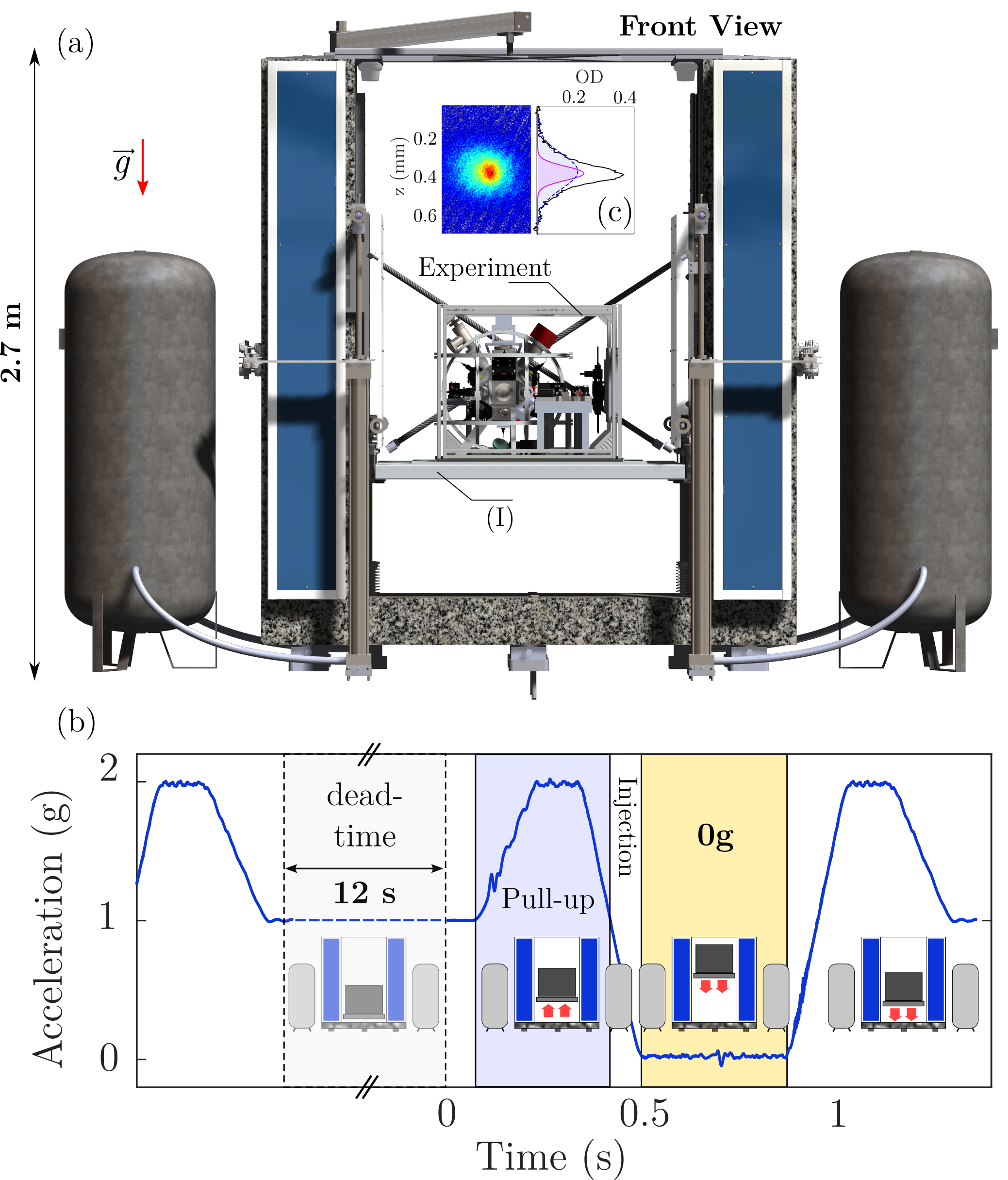}
  \caption{(a) Schematic of the Einstein elevator. The payload (I), including the science chamber, cooling beam optics and imaging system, is driven vertically along two air-bearing translation stages. (b) Acceleration profile of the science chamber for sequential parabolas of 400 ms, separated by a dead time of 12 s required to let the motors cool down. (c) Absorption image of the BEC transition after a time-of-flight of 50 ms in 0g. The projection of the vertical axis shows the typical double structure of the cloud. The blue (violet) curve is a least-squares fit to the thermal distribution (condensed fraction).}
  \label{fig:expsim}
\end{figure}

In this letter, we present an alternative method where $^{87}$Rb BECs are produced all-optically on an Einstein elevator, as shown in \Fig \ref{fig:expsim}. Our method relies on the combination of three optical techniques. First, we use a $\Lambda$-enhanced grey molasses on the $5^2$S$_{1/2} \rightarrow 5^2$P$_{3/2}$ D$_2$ transition to create a reservoir of cold atoms \cite{Rosi2018}. Second, our 1550 nm optical dipole trap (ODT) creates a transparency volume due to a strong light shift of the $5^2$P$_{3/2}$ excited state that results from coupling with the $5^2$D$_{5/2}$ state \cite{Clement2009}. In this way, it is possible to store the atoms in the dipole trap without emission and reabsorption of near-resonant photons \cite{Stellmer2013}. Third, we create a time-averaged or \emph{painted} potential by spatially modulating the ODT beam. This technique leads to both a high capture volume and fast evaporative cooling with a high collision rate \cite{Roy2016}. We produce a BEC in 1.4 s with a critical temperature of 140 nK before the Einstein elevator reaches the microgravity (0g) phase of its trajectory.

\begin{figure}[!bt]
  \centering
  \includegraphics[width=0.48\textwidth]{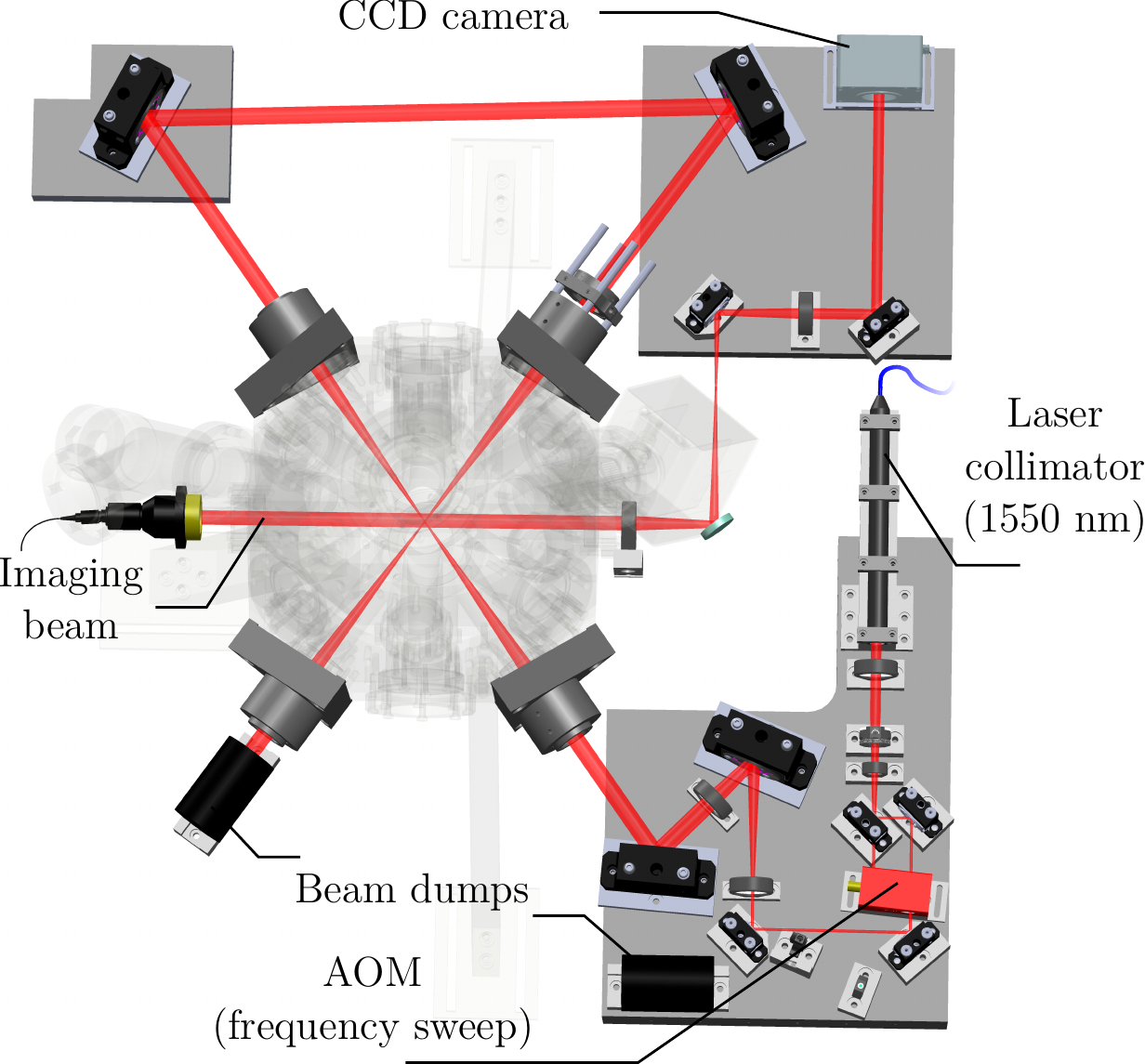}
  \caption{Details of the painted optical trap composed of two crossed beams and the imaging system.}
  \label{fig:DipoleTrapScheme}
\end{figure}

%------------------------ General Description ------------------------
Our experimental apparatus has been previously developed for parabolic flights onboard the Novespace Zero-G aircraft \cite{Barrett2014a, Barrett2016}. It comprises fiber-based lasers, an ultra-stable frequency source and a titanium science chamber housed inside a magnetic shield [\Fig \ref{fig:expsim}(a)]. Figure \ref{fig:DipoleTrapScheme} shows the optical layout of the crossed ODT beams, which are derived from a $\lambda_D = 1550$ nm amplified fiber laser delivering up to 23 W. After a first telescope, an acousto-optic modulator (AOM) is used to control the optical power and to spatially modulate the beams for the painted potential. A second telescope adapts the beam profile to obtain a beam waist of $w_0 = 45$ $\mu$m at the location of the atoms. The trap is formed by two beams crossed at an angle of $70^\circ$.
The key advantage of the painted potential is to break the fixed power-law relationship between the trap depth $U$ and the frequencies $\omega$ by modifying the shape of the trap. We restrict the discussion to pure harmonic traps with frequencies $\omega_y^2 = \omega_y^2 = \frac{8\alpha P}{\pi m w_0^4} f_\omega(h/w_0)$ and $\omega_z^2 = \frac{4\alpha P}{\pi m w_0^2 z_R^2} f_{\omega}(h/w_0)$, and depths $U = \frac{2\alpha P}{\pi w_0^4} f_U(h/w_0)$, where $\alpha$ is the polarizability of the atomic ground state at $\lambda_D$, $P$ is the optical power, $m$ is the atomic mass, $z_R = \pi w_{0}^{2}/\lambda_D$ is the Rayleigh length, and $h$ is the amplitude of the spatial modulation. The functions $f_{\omega}(h/w_0)$ and $f_U(h/w_0)$ are the fractional reduction factors due to the spatial modulation of the beam \cite{Roy2016}. The capture volume is proportional to $h^2$ and can be increased with a good trade-off on the trap depth $U$ due to the factor $f_U(h/w_0)$.

%------------------------ Loading of the trap ------------------------
We first load a 3D magneto-optical trap of approximately $5 \times 10^7$ $^{87}$Rb atoms from background vapour in 4 s. Atoms are cooled further by shifting the cooling beam detuning to $\Delta = -140$ MHz relative to the $|5^{2}$S$_{1/2},F=2\rangle \rightarrow |5^{2}$P$_{3/2},F'=3\rangle$ transition, while keeping the repumping beam near resonance with $|5^{2}$S$_{1/2},F=1\rangle \rightarrow |5^{2}$P$_{3/2},F'=2\rangle$. This repump frequency is derived from the cooling beam using a fibered electro-optic modulator (EOM) operating at 6.56 GHz. We then turn on the ODT with a spatial modulation amplitude of $h = 100$ $\mu$m. The frequency of the cooling beam is then shifted further to $\Delta = -240$ MHz, and we tune the repump to satisfy the Raman condition between $|F=1\rangle$ and $|F=2\rangle$ at a frequency difference of 6.834 GHz. This results in $\Lambda$-enhanced cooling and efficient ODT loading---combining the effects of grey molasses and velocity-selective coherent population trapping \cite{Rosi2018}.

Outside the dipole trap \footnote{This cooling process can also be implemented inside the dipole trap.}, grey-molasses cooling results from a spatially-varying light shift that allows moving atoms to undergo a Sisyphus-like cooling process until being optically pumped to a dark state $|{\rm DS}\rangle$. Atoms stay in $|\rm{DS}\rangle$ until velocity-induced motional coupling brings them back to a bright state where cooling begins again. The lifetime of $|{\rm DS}\rangle$, where atoms do not scatter light, sets the minimum achievable temperature. For the $|F = 2\rangle \rightarrow |F' = 2\rangle$ transition, only the ground states $|m_F = 0, p = 0\rangle$ and $|m_F = \pm 2, p = \pm 2\hbar k\rangle$ are coupled, where $m_F$ is a magnetic quantum number, $p$ is the momentum, and $k$ is the wavevector of the cooling light. Hence, the dark state can be written as the superposition of three momentum states $|{\rm DS}\rangle = \frac{1}{\sqrt{3}}(|-2,-2\hbar k\rangle - |0,0\rangle + |2,2\hbar k\rangle)$. As a result of this superposition, $|{\rm DS}\rangle$ is very short lived which limits the minimum temperature. In the $\Lambda$-enhanced scheme, the second laser is blue detuned from $|F = 1\rangle \rightarrow |F' = 1\rangle$, and is phase locked to the first via the EOM. Consequently, $|{\rm DS}\rangle$ will mix with states in the $F = 1$ manifold, forming a more complex but longer-lived non-coupled state $|{\rm NC}\rangle = (\Omega_H |G\rangle - \Omega_G |H\rangle)/ \sqrt{\Omega_G^2 + \Omega_H^2}$ where cold atoms are stored for larger durations---enhancing the cooling effect. Here, $|G\rangle$ and $|H\rangle$ represent superpositions of different magnetic sub-levels within the $F = 1$ and $F = 2$ manifolds, and $\Omega_{G,H}$ are effective optical pumping rates between these levels. The specific combinations of levels depends on the relative intensities of the cooling and repumping beams, and their polarization gradient potentials \cite{Grier2013}.

\begin{figure}[!tb]
  \centering
  \includegraphics[width=0.48\textwidth]{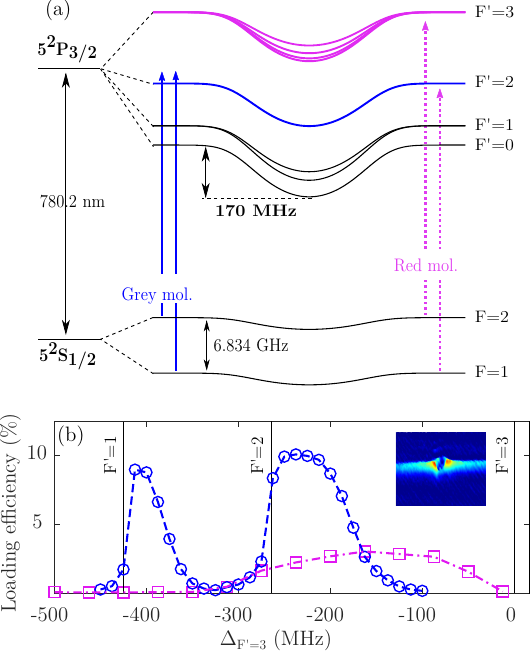}
  \caption{(a) Level structure of the atoms including the light shift due to the painted ODT ($P \simeq 10$ W, $h = 100$ $\mu$m). The cooling and repumping frequencies are shown for the grey (solid blue lines) and red (dashed pink lines) molasses. The light shift of the excited (ground) state at the center of the ODT is $\sim 170$ MHz (2.5 MHz). (b) ODT loading efficiency as a function of the detuning relative to $|F=2\rangle \rightarrow |F'=3\rangle$ for the red (pink squares) and grey (blue circles) molasses. Inset: absorption image of atoms in the ODT using near-resonant light. Atoms at the center of the trap are transparent to the cooling beams.}
  \label{fig:LoadingTrap}
\end{figure}

As atoms move toward the center of the ODT, the grey-molasses light becomes increasingly blue detuned from $|F=1,2\rangle \rightarrow |F'=2\rangle$, which strongly reduces light scattering. The total power in the ODT beams during this loading phase is 10 W. Under these conditions, the light shift of $|F'=2\rangle$ is $-170$ MHz at the trap center, as shown in \Fig \ref{fig:LoadingTrap}(a). This leaves the $|F=1,2\rangle \rightarrow |F'=3\rangle$ transitions far off resonance---enabling low temperature atoms to become trapped in the transparency volume of the ODT [see inset of \Fig \ref{fig:LoadingTrap}(b)]. In principle, the ODT-induced light shifts hinder grey molasses cooling \cite{Clement2009}. Nevertheless, in our painted potential, atoms still have a strong probability to be cooled and pumped into $|{\rm NC}\rangle$ because the pumping rate ($\sim 6$ MHz) is much higher than the modulation frequency of 280 kHz. Consequently a large fraction (about half) of the atoms should still undergo grey molasses cooling while being in the ODT volume. The sharp features in \Fig \ref{fig:LoadingTrap}(b) comply with this hypothesis. When the cooling beams are switched off, approximately $5 \times 10^6$ atoms remain in the painted potential with an in-trap temperature of 15 $\mu$K at a depth of 120 $\mu$K. We evaluate the gain of the grey molasses process by comparing it with a loading sequence using a standard red molasses. In this case, because of the light shifts, the optimal detuning varies strongly with the optical power of the ODT. In contrast, the grey molasses yields a 4-fold increase in atom number compared to the red molasses, as shown in \Fig \ref{fig:LoadingTrap}(b). Moreover, the grey molasses scheme is relatively insensitive to the ODT power, and it occurs on faster timescales ($< 1$ ms) than red molasses ($> 5$ ms) \cite{Rosi2018}, even though the final temperatures are comparable. This effect is consistent with our increased loading with the grey molasses: the longer times required for red molasses cooling lead to expansion of the atoms, which lowers the central density and therefore the loading efficiency.

After the molasses and loading phases, trapped atoms remaining in $|F=2\rangle$ are pumped into the $|F=1\rangle$ manifold using light resonant with $|F=2\rangle \rightarrow |F'=2\rangle$. The ODT is compressed by first increasing the laser power to 20 W in 50 ms with the modulation on. The modulation amplitude is then ramped down to zero over 150 ms while simultaneously decreasing the ODT power to 14 W. At this stage the trap depth is 390 $\rm{\mu}$K, the collision rate is 5000 s$^{-1}$, and the trap frequencies are $\omega_x = 2\pi \times 1375$ Hz, $\omega_y = 2\pi \times 1130$ Hz, and $\omega_z = 2\pi \times 780$ Hz. We then proceed to a forced evaporative cooling stage by decreasing the power by a factor 400 in 1.2 s using a sequence of three linear ramps with different slopes to take into account the reduction in collision rate. We verified the efficiency of this method by reaching a BEC for a critical temperature $T_{\rm c} = 200$ nK with $10^5$ atoms in standard gravity (1g).

\begin{figure}[!t]
  \centering
  \includegraphics[width=0.48\textwidth]{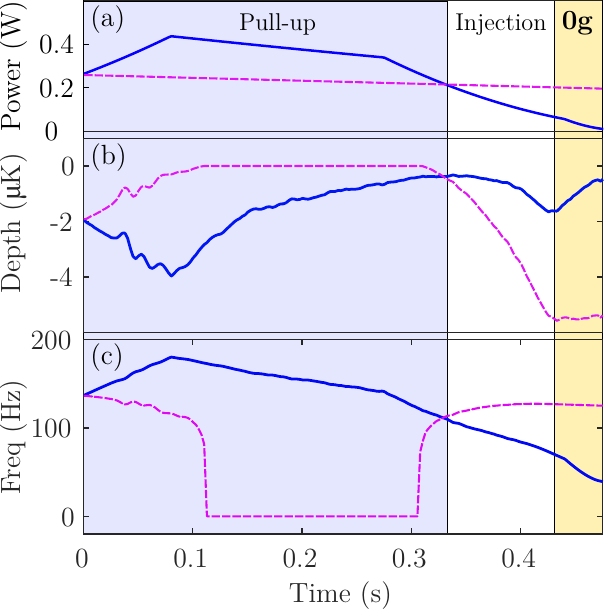}
  \caption{Adapted ODT parameters during the pull-up, injection and 0g phases (blue solid line). The parameters are calculated using the calibration of the ODT power and the measured acceleration. Calculations were validated by measuring the trap frequencies at three different times during the sequence: two in 1g and one in 0g. (a) ODT optical power. (b) Evolution of the ODT depth.  (c) Average ODT frequency taking into account the acceleration sag. For comparison, a sequence using parameters appropriate for 1g (pink dashed line) shows vanishing trap depth and frequency because of the increased sag during the pull-up phase.}
  \label{fig:FrequencyDepth}
\end{figure}

We realize BECs in microgravity by mounting our science chamber on an Einstein elevator developed by the French company Sym\'{e}trie (\Fig \ref{fig:expsim}), which undergoes pre-programmed parabolic trajectories that can provide up to 400 ms of weightlessness every 13.5 seconds. The residual acceleration during the motion, as measured with a low-noise mechanical accelerometer (Colibrys SF3600), yields a maximum amplitude of 1 m/s$^2$ and a root-mean-squared repeatability of 0.05 m/s$^2$ [see \Fig \ref{fig:expsim}(b)]. When operating on the elevator, the cooling sequence starts during the dead time between two parabolas, and the initial 800 ms of the evaporation process is the same as previously described. Just before the ``pull-up'' phase (when the acceleration increases above 1g), the total ODT power is $P = 250$ mW and the sample temperature is $T = 400$ nK. The evaporation sequence is optimized by experimentally maximizing the phase-space density at the end of the sequence. The varying acceleration during the pull-up phase strongly affects the trap depth and frequencies in time. To adapt to these changes, we apply a specific temporal profile to the ODT power (see \Fig \ref{fig:FrequencyDepth}). When the acceleration exceeds 1g, the trap is compressed adiabatically in 80 ms by increasing the power to $P = 426$ mW. This guarantees a depth sufficient (about 10 times the cloud temperature) to maintain the atoms in the trap when the acceleration reaches its maximum ($\sim 2$g). The evaporation is completed by ramping down the power in two steps until the end of the ``injection'' phase (when the acceleration decreases from 1g to 0g). Using this protocol, a BEC is obtained $\sim 100$ ms before the 0g phase, with $5 \times 10^4$ atoms at a critical temperature of $T_{\rm c} = 140$ nK. At this point, the trap frequencies are $\omega_x = 2\pi \times 109$ Hz, $\omega_y = 2\pi \times 103$ Hz, $\omega_z = 2\pi \times 71$ Hz. The total duration of the evaporation is 1.2 s.

\begin{figure}[!b]
  \centering
  \includegraphics[width=0.48\textwidth]{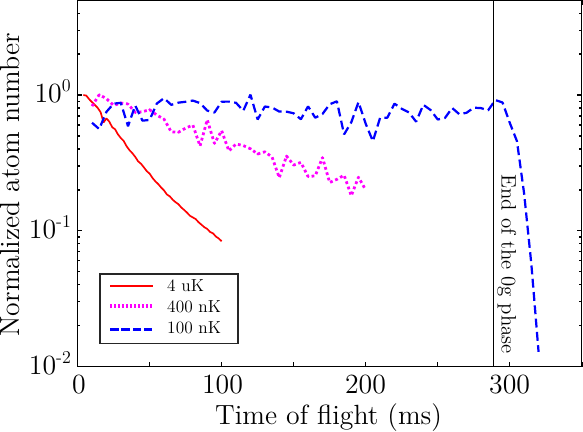}
  \caption{Normalized atom number detected versus the time of flight in 0g for three different sample temperatures.}
  \label{fig:zerogresults}
\end{figure}

At the beginning of the 0g phase, the ODT power is decreased in 40 ms to reach the minimum value required to keep the atoms in the trap. At this stage, our BEC contains $4 \times 10^4$ atoms for a spatial expansion corresponding to 35 nK. The ODT power is 10 mW for an average trap frequency of 39 Hz ($\omega_x = 2\pi \times 50$ Hz, $\omega_y = 2\pi \times 41$ Hz, $\omega_z = 2\pi \times 28$ Hz). We anticipate further improvements by implementing delta-kick collimation of the cloud \cite{Kovachy2015bis}. To illustrate the importance of reducing the temperature, we produced three samples with different temperatures and measured the fluorescence signal emitted from a small detection volume as a function of the time of flight in microgravity (see \Fig \ref{fig:zerogresults}). These data show that for $T \lesssim 100$ nK the atom number is constant inside the detection volume ($\sim 0.03$ cm$^3$) during the full time in free-fall. For larger temperatures, atoms expand outside the detection zone and the fluorescence signal decreases dramatically.

In conclusion we have produced all-optical BECs of $^{87}$Rb in microgravity with a cycle time of 13.5 s. The painted ODT, combined with light-shift engineering and $\Lambda$-enhanced grey molasses cooling, enabled us to optimally cool and evaporate atoms to quantum degeneracy before the 400 ms of weightlessness available for scientific studies. These techniques can easily be adapted to other atomic species, and are compliant with sympathetic cooling for multi-species experiments. Our Einstein elevator is an alternative solution to realize experiments in microgravity as currently proposed for the ISS to study few body systems \cite{Elliott2018}, new topological quantum objects such bubble-shaped traps \cite{Lundblad2019}, and ultra-cold atoms at extremely low temperatures \cite{Sackett2018}. Continuous operation of the elevator results in one hour of microgravity per day in a standard laboratory environment. This constitutes a new tool to prepare future Space-based missions considering ultra-cold atom interferometers for satellite gravimetry \cite{Douch2018}, quantum tests of the UFF \cite{Aguilera2014, Williams2016}, and gravitational wave detection \cite{Hogan2011}.

%================================================================================================
%\acknowledgments
%\section*{Acknowledgements}

This work is supported by the following agencies: CNES (Centre National d'Etudes Spatiales), l'Agence Nationale pour la Recherche and the D\'{e}l\'{e}gation G\'{e}n\'{e}rale de l'Armement under Grant ``HYBRIDQUANTA'' No. ANR-17-ASTR-0025-01, Grant ``TAIOL'' No. ANR-18-QUAN-00L5-02 and Grant ``EOSBECMR'' No. ANR-18-CE91-0003-01, the European Space Agency, IFRAF (Institut Francilien de Recherche sur les Atomes Froids), and the action sp\'{e}cifique GRAM (Gravitation, Relativit\'{e}, Astronomie et M\'{e}trologie). M.R. thanks CNES and IOGS for financial support. P.B. thanks Conseil R\'{e}gional d'Aquitaine for the Excellence Chair.

%================================================================================================
%\bibliographystyle{apsrev4-1}
\bibliography{References}
%================================================================================================

%%%%%%%%%%%%%%%%%%%%%%%%%%%%%%%%%%%%%%%%%%%%%%%%%%%%%%%%%%%%%%%%%%%%%%%%%%%%%%%%%%%%%%%%%%%%%%%%%
\end{document}